\documentclass[11pt,dvips]{article}
\usepackage{epsfig,times} 
\usepackage{picinpar}
\usepackage{wrapfig}
\usepackage{floatflt}
\makeatletter
\renewcommand{\section}{\@startsection{section}{1}{0in}
	{0.4\baselineskip}{0.1\baselineskip}{\Large\bf}}
\renewcommand{\subsection}{\@startsection{subsection}{2}{0in}
	{0.25\baselineskip}{-\baselineskip}{\large\bf}}
\renewcommand{\subsubsection}{\@startsection{subsubsection}{3}{0in}
	{0.1\baselineskip}{-\baselineskip}{\normalsize\bf}}
\makeatother
\begin{document}

%
\thispagestyle{myheadings}
%
\markright{OG 2.2.18}
\begin{center}
%
{\LARGE \bf Search for $\geq$ 400 GeV gamma-rays from the SNR Cas A\\
 with the CAT telescope}
\end{center}

\begin{center}
%
%
{\bf P. Goret$^{1}$, C. Gouiffes$^{1}$, E. Nuss $^{2}$} (for the CAT
collaboration) {\bf , and D.C.Ellison $^{3}$}\\
{\it $^{1}$DSM,DAPNIA/SAp, CE Saclay, Gif-sur-Yvette, 91191, Cedex France\\
$^{2}$}Groupe de Physique Fondamentale,Universite de Perpignan, Perpignan,
66860, France\\
$^{3}$Department of Physics, North Carolina State University, Raleigh,NC 27695, USA
\end{center}

\begin{center}
{\large \bf Abstract\\}
\end{center}
\vspace{-0.5ex}
%
%
The recent detection of a hard X-ray component in the supernova remnant 
Cassiopeia A
is interpreted as synchrotron emission from electrons accelerated to energies up
 to 40 TeV
(Allen et al., 1997). It is
therefore tempting to consider TeV gamma-ray emission from this object through :
i) bremsstrahlung and
inverse Compton radiation from electrons and/or ii) $\pi ^0$ production from an
associated high energy cosmic ray component hitting surrounding material. Cas A
was observed by the
CAT imaging Cherenkov telescope  during the 
observing season Aug-Nov 1998.
An upper limit to the integral flux above 400 GeV of 0.74$\times$10$^{-
11}$ $\gamma$cm$^{-2}$s$^{-1}$
is derived. This result is used to constrain shock-acceleration models for 
production of VHE gamma-rays in SNRs.

%

\vspace{1ex}

%
%
\section{Introduction:}
\label{intro.sec}
The supernova remnant Cassiopeia A (Cas A), aged $\sim$300 yr at a distance of $\sim$2.8 kpc, is the
brighest radio source in the sky. The radio emission region is concentrated in a shell of radius
$\sim$130 arcsec and thickness $\sim$30 arcsec. X-ray data from the RXTE satellite indicate the
presence of electrons accelerated to energies up to $\sim$40TeV which emit synchrotron photons
above 10 keV (Allen et al., 1997). 
These electrons could generate $\gamma$-rays in the GeV-TeV band via inverse
Compton or bremsstrahlung mechanisms. The $\gamma$-ray emissivity critically depends on the value
of the magnetic field in the acceleration region.
In addition, a nuclear component is likely to be
accelerated along with electrons giving rise to additional $\gamma$-rays 
produced in nuclear collisions via $\pi^{0}$ decay. 
Upper limits to the Cas A $\gamma$-ray flux 
above 100 MeV 
were obtained by the satellite experiments SAS-II and COS-B. Analyzing these data, Cowsik and Sarkar
(1980)  were able to set a lower limit of 8$\times$10$^{-5}$G to the magnetic field in Cas A.
Further observations by the EGRET experiment on the CGRO satellite led Esposito et al. (1996) to
derive a more constraining lower limit of 2$\times$10$^{-4}$G.

The situation in Cas A is reminiscent of that in SN1006 where Koyama et al.
(1995) inferred the presence of 
up to 100 TeV electrons from the observed X-ray
synchrotron emission. Subsequently, TeV $\gamma$-ray emission was observed by
the CANGAROO atmospheric Cherenkov telescope (Tanimori et al., 1998).
The Whipple collaboration, analyzing a 2.7 hr exposure on Cas A, 
reported an upper limit to the $\gamma$-ray flux
above 300 GeV of 2.6$\times$10$^{-11}$cm$^{-2}$s$^{-1}$ (Lessard, 1996).
More recently, the same group derived an improved upper limit of  
1.8$\times$10$^{-11}$cm$^{-2}$s$^{-1}$ above 400 GeV from a 6.9 hr exposure
(Lessard, 1999).
In this paper, we present an analysis of a 24.4 hr exposure on Cas A obtained
with the imaging atmospheric Cherenkov telescope CAT. The CAT telescope and the
data analysis technique are described in section 2. The database is presented
in section 3. The results of the analysis are summarized in section 4. The
implications for current models of $\gamma$-ray production in SNR's are
discussed in section 5.

\section{The CAT telescope and data analysis:}
\label{anal.sec}
The CAT imaging atmospheric Cherenkov telescope is described in Barrau et al.
(1998). It consists of a 18m$^2$ Davies-Cotton mirror fitted with a 600 PMT 
fine-grained camera (pixel size: 0.12$^\circ$ diameter) at the focal plane. 
The energy threshold is 250 GeV at zenith.
The telescope is located in the French Pyr\'en\'ees at an altitude of
1650m and geographical coordinates 2$^\circ$E, 42$^\circ$N. The data analysis
technique used in the present paper is detailed in Le Bohec et al. (1998). The
basis for the analysis of individual shower images is to compare the observed
image to a template of calculated images at different energies, 
zenith angles
and impact parameters to the telescope. Gamma/hadron discrimination is performed
with a cut on a single $\chi^{2}$-like parameter. The analysis also yields the
energy, impact parameter and orientation 
(the equivalent of the $\alpha$-parameter in 
the standard Hillas method) for each event. Note that, in the present study, Cas
A is considered as a point-like source, just as the Crab nebula is.
Extensive Monte-Carlo simulations of
the instrument have been performed and proved highly reliable when checked 
against real $\gamma$-rays from the Crab or Mkn501 (e.g. Goret, 1997, Mohanty,
1999).

\section{The Cas A database:}
\label{data.sec}
The observations relevant for the present analysis were performed during the
period August-November 1998. ON-source runs lasted for 30 minutes followed or
preceded (but not systematically) by OFF-source runs shifted by $\pm$ 35 minutes
in RA. After selection on meteorological conditions, 
zenith angle ($\leq$25$^\circ$) and correct experiment operation, a total 
exposure of 24.4 hr ON-source and 13.6 hr OFF-source were retained for further
analysis. A database on the Crab nebula, with the same selection criteria and 
observing period, was also compiled to allow for cross-checking of the analysis. 
The trigger condition required that 4 PMTs above a threshold of 2.5
photoelectrons fire simultaneously within a sector 
(see Barrau et al., (1998), for details).
The average raw trigger rate for these two data samples was $\sim$17 Hz. Both
the Cas A and Crab data were processed the same way as is decribed in the next
section.

\begin{figure}[htb]
\centering
\mbox{\epsfig{figure=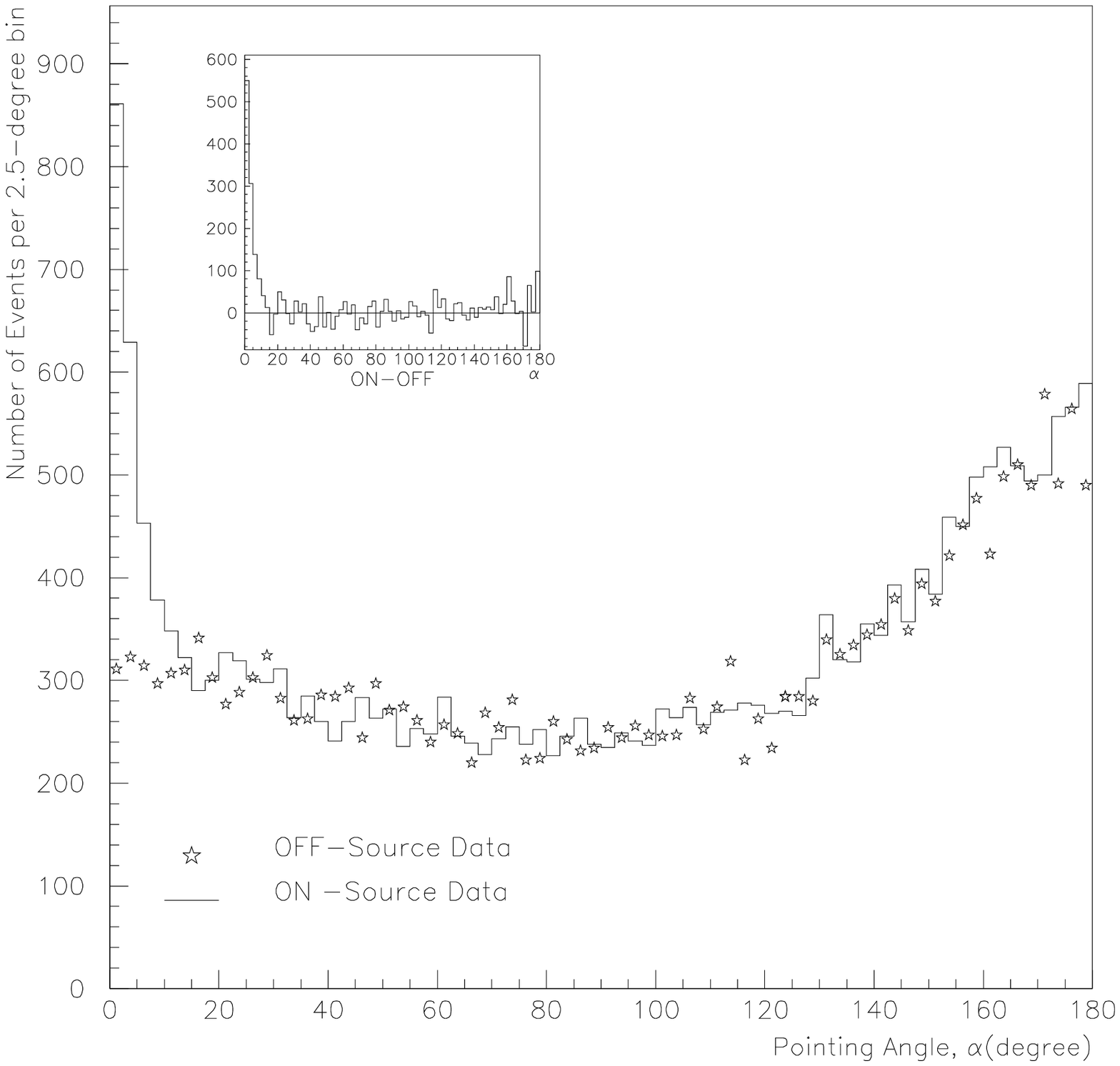,height=8.7cm}
      \epsfig{figure=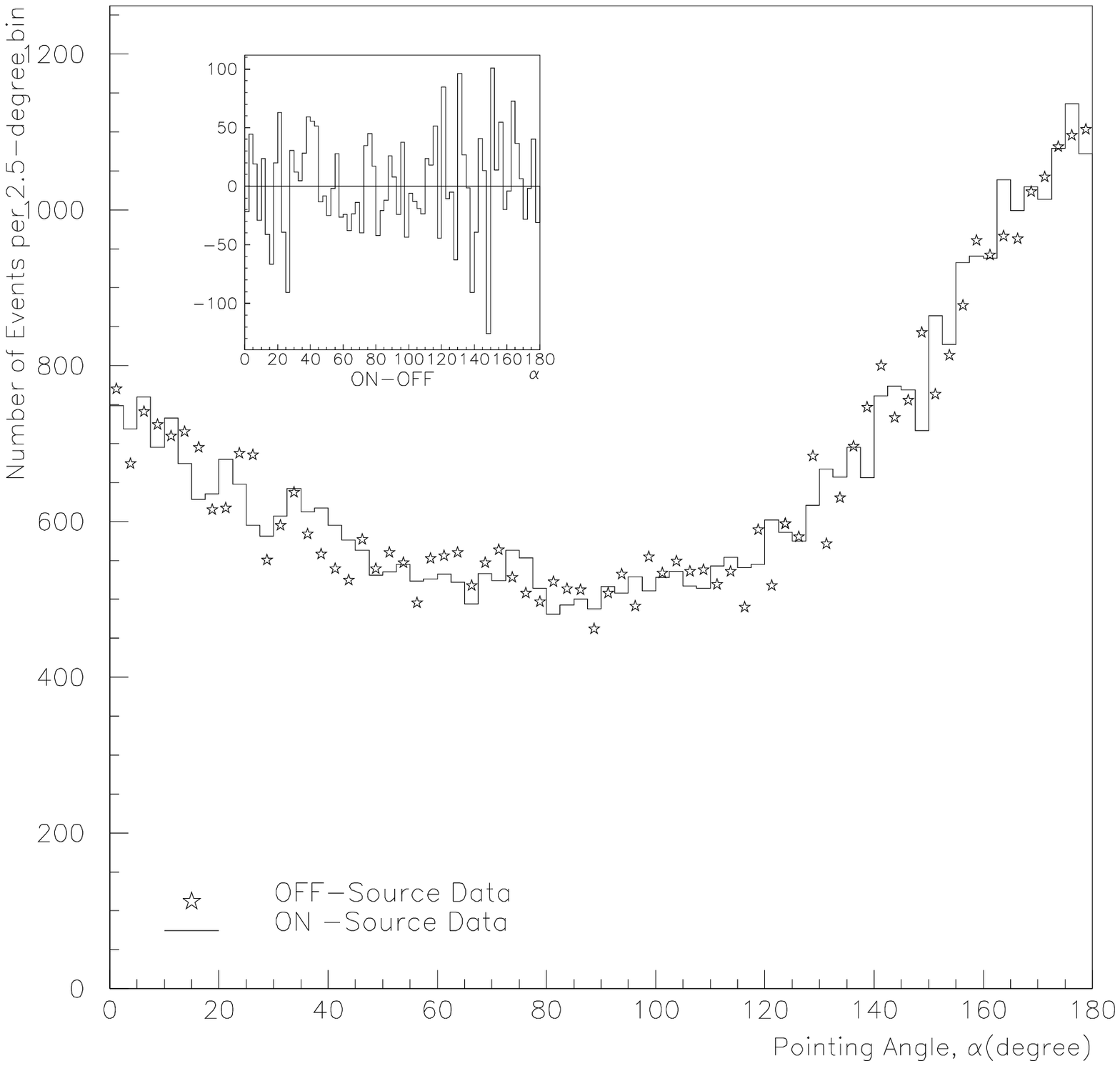,height=8.7cm}}
\vspace*{-2.5cm}\\
\hspace*{6.0cm}{\bf a)}\hspace*{8.5cm}{\bf b)}\vspace*{1.5cm}
\caption{\it The distributions of the orientation angle $\alpha$ for 
the Crab (a) and Cas A (b) datasets respectively using the
$\chi^2$ method. The ON and (renormalized) OFF-source data are 
represented by the solid
histogram and stars respectively. 
The insets shows the ON-OFF distributions.
The data are binned in 2.5$^\circ$ intervals.}
\end{figure}  

\section{Results:}
\label{results.sec}
The Crab data were used to optimize the cuts for the best sensitivity in
the search for a signal from Cas A. This analysis yielded the following cuts for
gamma/hadron (shape) discrimination: i) $\chi^{2}$ probability $\geq$0.40 and 
ii) a total number of
photoelectrons in the image of Q$_\mathrm{tot}$$\geq$50 p.e.. In addition, the estimated
impact parameter was restricted to $\leq$160 m. 
The $\alpha$-plots for the Crab and Cas A are displayed in figure 1. 
The OFF data were normalized using the tracking ratio, i.e., the
ratio n$_\mathrm{ON}$/n$_\mathrm{OFF}$ of the number of events surviving the shape cuts 
in the region 20$^\circ \leq \alpha \leq$120$^\circ$.
The orientation cut was set to
$\alpha _{max}$=8$^\circ$, which 
retains $\sim$ 80\% of the $\gamma$-rays selected after the shape cuts
according to Monte-Carlo simulations. 
The results of the analysis are summarized in Table 1.

\vskip8truept
\begin{center}
\begin{tabular}{|llll|}\hline
& & & \\
& Crab & & Cas A \\
& & & \\
~~~T$_\mathrm{ON}$ (hrs) & 13.3 & \hspace*{1cm}  & 24.4~~~ \\
~~~T$_\mathrm{OFF}$ (hrs) & 9.2 &   & 13.6~~~ \\
~~~N$_\mathrm{ON}$ (events) & 2006 &   & 2360~~~ \\
~~~N$_\mathrm{OFF}$ (events) & 706 &   & 1259~~~ \\
~~~Tracking ratio & 1.43 & & 1.85~~~ \\
~~~Excess (events) & 998 &   & 33.5~~~ \\
~~~Standard Deviation (events) \hspace*{1cm}& 60.7 &   & 85.9~~~ \\
~~~Significance (n$_{\sigma}$) & 16.45 &   & 0.39~~~ \\ 
& & & \\
\hline
\end{tabular}
\end{center}

\centerline{Table 1. {\it Results of the analysis for the Crab and Cas A.}}

\vskip8truept
>From these data, a 3$\sigma$ upper limit to the Cas A flux can be set at
$\leq$0.14 times the Crab flux.
Monte-Carlo calculations indicate that the energy threshold subsequent to the
above cuts is 400 GeV, with $\sim$89\% of the selected events in the
sample having an effective energy above 400 GeV. The effective area,
assuming a Crab-like energy spectrum for Cas A, is estimated to be
$\sim$3.5$\times$10$^{8}$cm$^{2}$. With the assumption of a null detected signal,
this translates into a 3$\sigma$ upper limit
to the Cas A $\gamma$-ray flux of:
\begin{center}
$\Phi$($\geq$400 GeV) $\leq$ 0.74$\times$10$^{-11} \gamma$cm$^{-2}$s$^{-1}$
\end{center}
The implications of this result concerning the current models for TeV-gamma
production in SNRs are examined in the next section.

\begin{figure}[htb]
\centerline{\epsfig{file=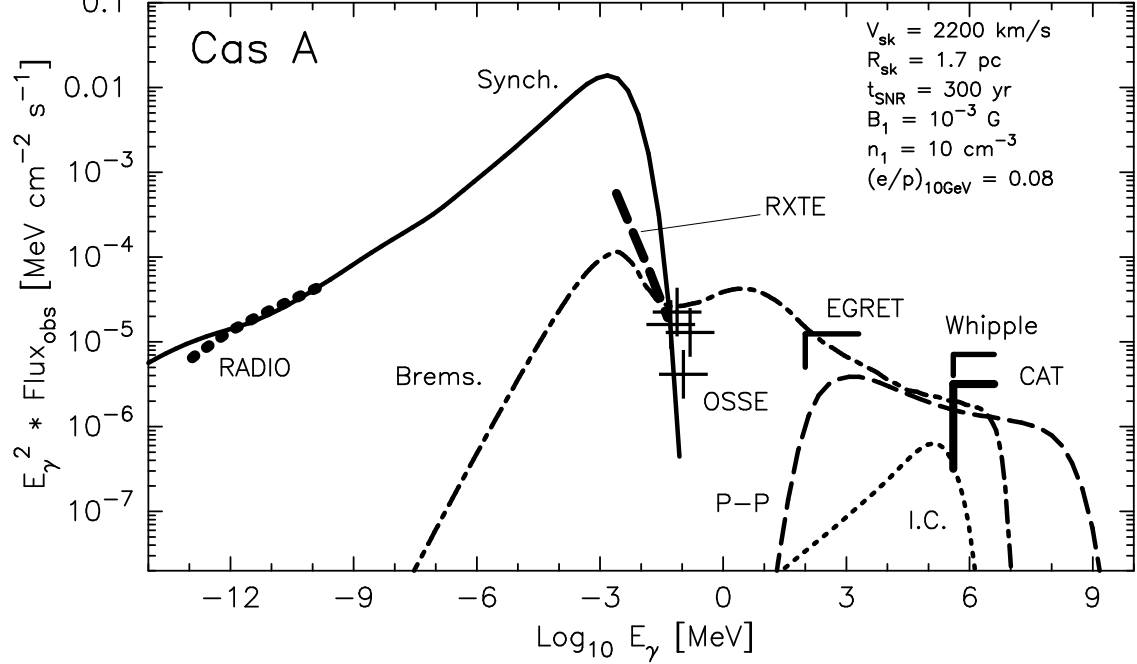,height=9.0cm}}
\caption{\it Predictions of the nonlinear diffusive shock-acceleration model from radio 
to TeV
$\gamma$-rays as compared to observations (see Ellison et al. (1999) for details and
references therein). The present result is shown together with the Whipple upper limit of
Lessard (1999).}
\end{figure}

\section{Discussion:}
\label{discuss.sec}
As was already noted by Cowsik and Sarkar (1980), considering $\gamma$-ray
production via bremsstralhung and inverse Compton, upper limits on the
$\gamma$-ray flux from Cas A must constrain primarily the strength of the
magnetic field in the source. The problem was recently reconsidered by Baring et
al. (1999) using a non-linear shock-acceleration model. The model aims at
predicting the overall continuum photon emission spectrum in SNRs 
from radio to $\gamma$-rays. The model was tailored to the specific case of Cas
A by Ellison et al (1999), taking into account relevant observations over the
full photon energy range. The results are displayed in figure 2, which shows the
best fit of the model to the data from radio through X-rays up to
$\gamma$-rays. The contributions of the different components (synchrotron,
bremsstrahlung, IC and p-p) in each energy band are indicated. The conclusion is
that in order to account for the upper limits reported in both the GeV and TeV
ranges, the strength of the magnetic field must be greater than
$\sim$10$^{-3}$G. As discussed by Cowsik and Sarkar (1980), such a high value in
excess of the equipartition field of $\sim$4$\times$10$^{-4}$G calls for a
magnetohydrodynamic field amplification which remains to be 
fully investigated (see Keohane, (1998), for a recent discussion). The
present result may also be considered in the frame of the diffusive 
shock-acceleration model of Drury, Aharonian and V\" olk (1994) as a test of
cosmic-ray acceleration in SNRs.

\section{Conclusions:}
\label{concl.sec}
The improved upper limit to the TeV $\gamma$-ray flux from the SNR Cas A
presented in this paper sets a stringent lower limit to the magnetic field in
the acceleration region. This result should help understanding the
mechanisms at work in this young object.

%
%

%
%
%
%
%
%
\vspace{1ex}
\begin{center}
{\Large\bf References}
\end{center}
%
Allen, G.E., et al. 1997, ApJ 487, L97 \\
Baring, M.G., et al. 1999, ApJ 513, 311 \\
Barrau, A., et al. 1998, Nucl. Instr. Meth. A 416, 278 \\
Cowsik, R., \& Sarkar, S. 1980, MNRAS 191, 855 \\
Drury, L.O'C. et al. 1994, A\& A 287, 959 \\
Ellison, D.C., et al. 1999, this conference OG 2.2.09 \\
Goret, P. 1997, Proc. Kruger TeV Workshop on Gamma-Ray Astrophysics, p.166 \\
Keohane, J.W. 1998, PhD thesis, University of Minnesota \\
Koyama, K., et al. 1995, Nature 378, 255 \\
Le Bohec, S., 1998, Nucl. Instr. Meth. A 416, 425 \\
Lessard, R. 1996, PhD thesis, Purdue University \\
Lessard, R. 1999, Proc. 19th Texas Symposium, Paris 1998, in press \\
Mohanty, G. 1999, this conference OG 2.2.03 \\
Tanimori, T., et al. 1998, ApJ 497, L25 \\

%

\end{document}